\documentclass[3p]{elsarticle}

\usepackage{lineno, hyperref}

\journal{}

\usepackage{multicol}
\usepackage{color}
\usepackage{xcolor}
\usepackage{float}
\usepackage{dcolumn,booktabs}
\usepackage{hyperref}
\newcolumntype{d}[1]{D{.}{.}{#1}}
\usepackage{outlines}
\usepackage{nomencl}
\usepackage{multirow}
\usepackage{soul}
\usepackage{marginnote} 
\usepackage{amssymb} 
\usepackage{amsmath}
\usepackage{subcaption} 
\usepackage{algorithm}
\usepackage{algpseudocode}

\makenomenclature
\usepackage{etoolbox}
\renewcommand\nomgroup[1]{%
  \item[\bfseries
  \ifstrequal{#1}{P}{{Parameters and variables}}{%
  \ifstrequal{#1}{S}{{Subscripts}}{%
  \ifstrequal{#1}{T}{{Acronyms}}{}}}%
]}
\RequirePackage{ifthen}
\newcommand{\nomunit}[1]{%
\renewcommand{\nomentryend}{\hspace*{\fill}#1}}

\begin{document}

\begin{frontmatter}

\title{Market-Oriented Flow Allocation for Thermal Solar Plants: An Auction-Based Methodology with Artificial Intelligence}

\author[add]{Sara Ruiz-Moreno\corref{cor1}}\ead{srmoreno@us.es}
 \author[add]{Antonio J. Gallego}
 \ead{agallego2@us.es}
 \author[add2]{Manuel Macías}\ead{manuel.macias@atlanticayield.com}
 \author[add3]{Eduardo F. Camacho}\ead{efcamacho@us.es}
 \cortext[cor1]{Corresponding author}

\address[add]{Dept. de Ingenier\'{i}a de Sistemas y Autom\'{a}tica, University of Seville, Camino de los Descubrimientos, no number, E-41092, Seville, Spain}
\address[add2]{Atlantica Yield, Albert Einstein s/n, 41092 Seville, Spain}
\address[add3]{AICIA - Dept. de Ingenier\'{i}a de Sistemas y Autom\'{a}tica, University of Seville, Camino de los Descubrimientos, no number, E-41092, Seville, Spain}

\begin{abstract}

This paper presents a novel method to optimize thermal balance in parabolic trough collector~(PTC) plants. It uses a market-based system to distribute flow among loops combined with an artificial neural network~(ANN) to reduce computation and data requirements. This auction-based approach balances loop temperatures, accommodating varying thermal losses and collector efficiencies. Validation across different thermal losses, optical efficiencies, and irradiance conditions—sunny, partially cloudy, and cloudy—show improved thermal power output and intercept factors compared to a no-allocation system. It demonstrates scalability and practicality for large solar thermal plants, enhancing overall performance. The method was first validated through simulations on a realistic solar plant model, then adapted and successfully tested in a 50 MW solar trough plant, demonstrating its advantages. Furthermore, the algorithms have been implemented, commissioned, and are currently operating in 13 commercial solar trough plants.

\end{abstract}


\end{frontmatter}


\nomenclature[P]{$G$}{{Collector aperture}\nomunit{m}}
\nomenclature[P]{$H_{\text{t}}(T)$}{{Convective heat transfer coefficient}\nomunit{W{\slash} ($\mathrm{m}^2$ $^\circ$C)}}
\nomenclature[P]{$A$}{{Pipe cross-sectional Area}\nomunit{$\mathrm{m}^2$}}
\nomenclature[P]{$\delta_\text{s}$}{{{Declination}}\nomunit{{$^\circ$}}} 
\nomenclature[P]{$\rho(T)$}{{Density}\nomunit{kg\slash$\mathrm{m}^3$}}
\nomenclature[P]{$I(t)$}{{Direct solar irradiance}\nomunit{W\slash$ \mathrm{m}^2$}}
\nomenclature[P]{$q$}{{Flow rate}\nomunit{m$^3$\slash s}} 
\nomenclature[P]{$n_\text{o}(t)$}{{Geometric efficiency}\nomunit{$-$}}
\nomenclature[P]{$\omega_\text{s}$(t)}{{{Hourly angle}}\nomunit{{$^\circ$}}}
\nomenclature[P]{$\phi$}{{Latitude}\nomunit{${º}$}}
\nomenclature[P]{$L_\text{loop}$}{{Loop length}\nomunit{m}}
\nomenclature[P]{$N_\text{loops}$}{{Number of loops}\nomunit{-}}
\nomenclature[P]{$N_{\text{it}}$}{{Number of auction iterations}\nomunit{$-$}}
\nomenclature[P]{$N_{\text{it,v}}$}{{Number of valve aperture iterations}\nomunit{$-$}}
\nomenclature[P]{$K_{\text{opt}}$}{{Optical efficiency}\nomunit{$-$}}
\nomenclature[P]{$\alpha_{K_{\text{opt}}}$}{{Optical efficiency fault}\nomunit{$-$}}
\nomenclature[P]{$C(T)$}{{Specific heat capacity}\nomunit{J{\slash}  (kg$^\circ$C)}}
\nomenclature[P]{$P_\text{cp}$}{{Specific heat capacity per unit volume}\nomunit{J\slash $^{\circ}$C m$^3$}}

\nomenclature[P]{$T(t,{x})$}{{Temperature}\nomunit{$^\circ$C}}
\nomenclature[P]{$H_\text{l}(T)$}{{Thermal loss coefficient}\nomunit{W{\slash}($\mathrm{m}^2$$^\circ$C)}}
\nomenclature[P]{$t$}{{Time}\nomunit{s}}
\nomenclature[P]{$S$}{{Total area of the field}\nomunit{m$^2$}}
\nomenclature[P]{$L$}{{Tube perimeter}\nomunit{m}}
\nomenclature[P]{$q^{+}$}{{Increased flow rate}\nomunit{m$^3$\slash s}}
\nomenclature[P]{$q^{-}$}{{Decreased flow rate}\nomunit{m$^3$\slash s}}
\nomenclature[P]{$\Delta q$}{{Quantum of flow rate}\nomunit{m$^3$\slash s}}
\nomenclature[P]{$P$}{{Power}\nomunit{W}}
\nomenclature[P]{$P^{+}$}{{Power after increasing flow rate}\nomunit{W}}
\nomenclature[P]{$P^{-}$}{{Power after decreasing flow rate}\nomunit{W}}
\nomenclature[P]{$P_{\text{demand}}$}{{Demand power}\nomunit{W}}
\nomenclature[P]{$P_{\text{supply}}$}{{Supply power}\nomunit{W}}
\nomenclature[P]{$C_{\text{demand}}$}{{Demand price}\nomunit{kg\slash s$^2$}}
\nomenclature[P]{$C_{\text{supply}}$}{{Supply price}\nomunit{kg\slash s$^2$}}
\nomenclature[P]{$P_{\text{au}}$}{{Auction power}\nomunit{W}}
\nomenclature[P]{$Q$}{{Sector flow rate}\nomunit{m$^3$\slash s}}
\nomenclature[P]{$v$}{{Valve aperture}\nomunit{-}}
\nomenclature[P]{$t_{\text{s}1}$}{{External controller sample time}\nomunit{s}}
\nomenclature[P]{$t_{\text{s}2}$}{{Apertures sample time}\nomunit{s}}
\nomenclature[P]{$\bar{T}$}{{Average temperature}\nomunit{$^\text{o}$C}}
\nomenclature[P]{$\bar{IF}$}{{Average intercept factor}\nomunit{-}}
\nomenclature[P]{$\delta$}{{Scaling factor for the supply and demand prices}\nomunit{$-$}}
\nomenclature[P]{$k$}{{Thermal power penalty factor}\nomunit{$-$}}

\nomenclature[S]{a}{Ambient}
\nomenclature[S]{f}{Fluid}
\nomenclature[S]{in}{Input}
\nomenclature[S]{mean}{Mean between input and output}
\nomenclature[S]{out}{Output}
\nomenclature[S]{max}{Maximum}
\nomenclature[S]{th}{{Thermal}}

\setlength\marginparsep{-550pt} 
\setlength\marginparsep{5pt} 

\section{Introduction}

The Sun is the primary energy source, underpinning nearly all other forms of energy, both fossil-based and renewable~\cite{solarTrends}. Its immense potential for harnessing energy and its environmentally sustainable characteristics position solar power as an increasingly compelling solution to meet the world's growing energy demands~\cite{PVsurvey}. In fact, solar and wind power have grown from contributing less than 2\% to 12\% of global electricity generation since 2010~\cite{wec_humanising}.

Solar energy is typically harnessed through two main technologies: photovoltaics (PV) and concentrating solar power (CSP). Among the various CSP technologies, parabolic trough collectors (PTCs) stand out because of their numerous advantages, including high efficiency, a low environmental footprint, and seamless integration with existing energy infrastructure. Additionally, their scalability-from small systems for individual buildings to expansive commercial plants covering several hectares-makes them particularly advantageous~\cite{ahmad2023PTC, NMPCbalance}.

In PTC plants, solar irradiance is typically measured by a single pyrheliometer~\cite{pyrheliometer} for the entire plant or one per sector. Given the vast size of commercial plants, this layout cannot provide detailed information about clouds covering only a few collectors. This leads to uncertainties in the exact irradiance at each collector. Moreover, reflectivity varies across loops due to factors like dust accumulation or breakage, resulting in imbalances across the field. To mitigate these effects, commercial plants often employ mechanisms such as defocusing {to protect against overheating issues and manipulate the input valves to improve the thermal balance~\cite{song2024homoKalman}.} Opening the valves of the most efficient loops helps reduce energy losses and balance the heat transfer fluid (HTF)~\cite{camacho2024control}.

Several studies have addressed the challenge of achieving thermal balance in PTC plants. Sánchez et al. ~\cite{TempHomo} developed an optimization algorithm to adjust the input valves of the loops in the ACUREX field using a centralized solution approach. They later refined this control approach~\cite{ThermalBalance} and applied it to a 50-MW large-scale PTC plant. The algorithm optimized valve apertures and was compared to a scenario where the input valves remained fixed. Given the large number of loops in commercial, large-scale plants, solving the optimization algorithm required substantial computational resources. To mitigate this issue, they clustered loops with similar efficiencies to reduce the number of decision variables. Later, Gallego et al.~\cite{gallego2023tcp100} applied temperature homogenization to the TCP-100 plant, incorporating a heuristic-based algorithm to handle strong transients affecting the field. Frejo and Camacho~\cite{frejo2020CentralDistributedMPC} proposed another strategy to reduce computational time by designing a distributed model predictive control (MPC) algorithm for calculating input valve apertures. Their approach achieved performance comparable to the centralized method but with lower computational effort. Chanfreut et al. ~\cite{chanfreut2023clustering} introduced a clustering-based method for controlling flow rates using a simplified plant model. Additionally, Sánchez et al.~\cite{sanchez2023populationDynamics} proposed a coalitional MPC approach to address the shared-resource constraint that couples local optimization problems. They introduced a population-dynamics-assisted resource allocation strategy to decouple these problems effectively.

Another approach to reducing computation time is leveraging artificial intelligence (AI), which can be applied to various components of the control system, from modeling~\cite{schimperna2024rnnmpc} to directly determining control actions~\cite{hose2023approxMPC}. In the context of thermal solar plants, several implementations are documented in the literature. For example, Cervantes-Bobadilla et al.~\cite{ANNi_PTC} combined an inverse artificial neural network (ANN) with particle swarm optimization to determine the optimal flow rate for achieving a target temperature. Goel et al.~\cite{goel2021genetic} employed a multi-objective genetic algorithm in a PTC plant to optimize the flow rate for a desired temperature. Tilahun~\cite{tilahun2024fuzzy} introduced a hybrid fuzzy convolution model for a PTC plant within a deep deterministic policy gradient algorithm, implemented in a fuzzy-based predictive deep reinforcement learning framework. Another solution involved integrating a coalitional MPC approach with neural networks trained to solve optimization problems~\cite{masero2022marketMPC}, achieving significantly reduced computation times with minimal performance loss in a 100-loop PTC plant.

The drawback of the latter methodology is that it relies on information that is not {usually} available in many plants, such as the spatial distribution of solar irradiance, the temperatures of the metal, {the thermal losses coefficient, etc. Furthermore, sensors are costly and difficult to calibrate. For this reason, commercial plants typically {have one or a small set of fully instrumented loops with more sensors than the rest of the plant. In these loops, all the temperature measurements, optical efficiency, and thermal losses are known or well-estimated. These values are extrapolated to the rest of the entire plant.} Additionally, variables such as metal temperature are not typically measured. 

{To apply these strategies to commercial solar trough plants, using only the typically available information in real plants would be essential.} This paper proposes a method to distribute and balance the temperatures of the loops using the information available in the plant. The methodology is based on a market system wherein each loop adjusts its flow rate based on an auction price. Furthermore, an ANN is employed to reduce computational costs and make the methodology implementable in the plant.  The method was adapted to a real 50 MW plant. The main contributions are:

\begin{itemize}
    \item Distribution of flow among loops using {only} the information available in the plant. 
    \item Auction-based mechanism with a market price determined by thermal power.
    \item Artificial neural network that learns the flow distribution and mimics the {auction}-based methodology. 
    \item Successful implementation and deployment of the algorithms in commercial solar power plants.
\end{itemize}

The paper is organized as follows: Section 2 provides a description of the system used, Section 3 presents the proposed methodology along with the defocusing mechanism, Section 4 presents the results obtained, and finally, Section 5 concludes the paper and suggests future lines of development.


\printnomenclature 

\section{System Description}

A PTC plant is a solar thermal system composed of loops of parabolic mirrors that focus sunlight onto a central focal line. A fluid, such as water or oil, flows through a pipe along this line, where it is heated to produce thermal energy. The HTF is typically directed to a steam generator, and it drives a turbine, as shown in Figure \ref{fig:PTC_plant_generic}.

\begin{figure*}[ht]
    \centering
    \includegraphics[width = \textwidth]{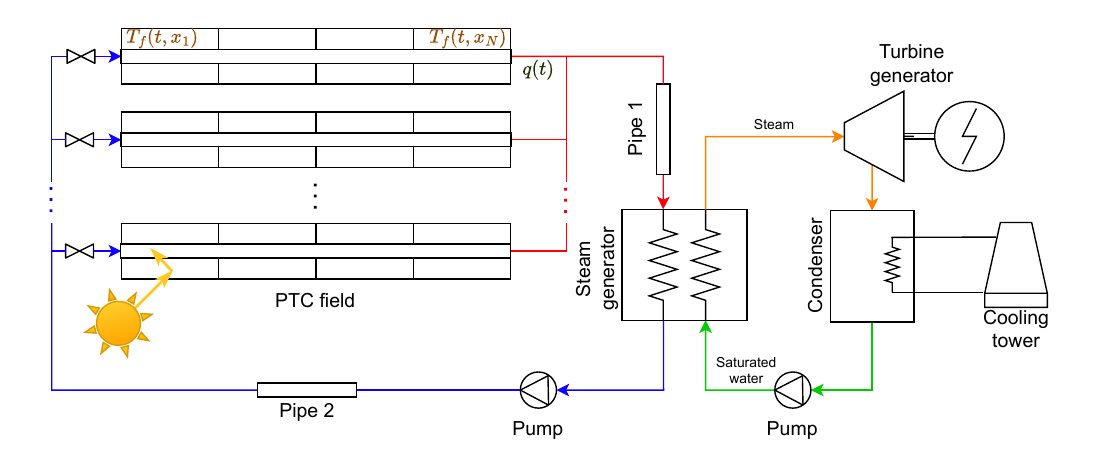}
    \caption{General scheme of a PTC plant.}
    \label{fig:PTC_plant_generic}
\end{figure*}

The simulations in this study were conducted using a generic 50 MW plant with characteristics similar to those of the Mojave plant~\cite{modelMojave}. Each loop has a total length of 620 m, with an active section of 593 m that receives solar irradiance. Each loop contains 4 collectors aligned in a north-south orientation. This analysis considered a sector consisting of ten loops ($N_\text{loops}=10$). The HTF used is Therminol VP-1~\cite{therminolvp1}, with a nominal operating temperature of around 390-393 $^\text{o}$C. Its density, $\rho_\text{f}$, and specific heat capacity, $C_\text{f}$, are defined by equations \ref{eq:htfrho} and \ref{eq:htfc}~\cite{MPCmojave}.

\begin{equation}\label{eq:htfrho}
    \rho_\text{f} = 1061.5-0.5787T_\text{f}-9.0242\cdot10^{-4}T_\text{f}^2
\end{equation}

\begin{equation}\label{eq:htfc}
    C_\text{f} = 1552.049+2.38501T_\text{f}+ 0.0010558T_\text{f}^2
\end{equation}

The geometric efficiency~\cite{princsolar,modelTCP100}, denoted as $n_\text{o}$ or $\cos(\theta)$, is determined by the correlation between the direction of the radiation beam and the mirror's perpendicular vector. This factor depends on several variables: collector dimensions, solar hour, hourly angle, declination, latitude, and Julian day. Since this plant is aligned in a north-south orientation, the geometric efficiency is calculated using Equation~\ref{eq:geometric_efficiency}~\cite{dynamicModelPTC}.

\begin{equation}
    \label{eq:geometric_efficiency}
\begin{aligned}
    n_\text{o} = \left( (\sin(\phi)\sin(\delta_\text{s})+\cos^2(\delta_\text{s})\sin^2(\omega_\text{s}) \right.  \\ + \left.  \cos(\phi)\cos(\delta_\text{s})\cos(\omega_\text{s}))^2     \right) ^{\frac{1}{2}}
\end{aligned}
\end{equation}

The plant features a sun-tracking system that precisely controls the rotation of the mirrors around an axis parallel to the pipe, optimizing the geometric efficiency for capturing and utilizing solar radiation~\cite{ControlSolar}.

The methodology applied in this work was tested on two different plant models. First, the auction-based mechanism was designed using the concentrated parameter model with a simple defocusing strategy to allow for a sufficiently fast implementation. The neural networks were trained using data obtained from simulations of this model {to allow fast ANN training due to its simplicity but without losing good temperature approximation}. Next, the resulting neural networks were applied to the distributed-parameter model to evaluate the robustness of the methodology, providing a more complex but slower representation of the system and a more realistic defocusing mechanism.

\subsection{Concentrated-Parameter Model}

The concentrated parameter model, also known as the lumped parameter model, provides a simplified representation of the plant by describing the variation in the internal energy of the fluid. This model is expressed by Equation \ref{eq:concentrated}. The thermal capacity of the loop is given by $C_\text{loop} = L_\text{loop} \rho_\text{f} C_\text{f} A_\text{f}$, while the specific heat capacity per unit volume is $P_\text{cp} = \rho_\text{f} C_\text{f}$. The multiplier $\alpha_{K_\text{opt}}$ accounts for variations in optical efficiency due to factors such as clouds, coating, dirt, degradation, breakage, and corrosion.
 
\begin{equation}
    \label{eq:concentrated}
\begin{aligned}
    C_\text{loop} \frac{d T_\text{out}}{d t} = (2-\alpha_{H_\text{l}})H_\text{l} A (T_\text{a} - T_\text{mean})   \\ + \alpha_{K_\text{opt}} n_\text{o} K_\text{opt} I S + q P_\text{cp} (T_\text{in} - T_\text{out})
\end{aligned}
\end{equation}

\noindent where $S = 3415.5$ m$^2$ and {the thermal loss coefficient, $H_\text{l}$, is defined by Equation~\ref{eq:Hl}.}

\begin{equation}\label{eq:Hl}
    \begin{split}
        & H_\text{l} =  1.137 \cdot 10^{-8} \left( T_\text{f} - T_\text{a} \right)^3 
        -  3.235 \cdot 10^{-6}  \left( T_\text{f} - T_\text{a} \right)^2 \\
        & +  1.444 \cdot 10^{-4}  \left( T_\text{f} - T_\text{a} \right) 
        + 8.179 \cdot 10^{-2} 
        - \frac{4.796}{T_\text{f} - T_\text{a}}
    \end{split}
\end{equation}

The initial tests for evaluating the methodology were conducted on the static version of the concentrated parameter model, which was derived by approximating the system by canceling the derivatives in Equation \ref{eq:concentrated}, resulting in Equation \ref{eq:static}. {This approach is fundamental because for the methodology to be easily extrapolated to an actual plant, the training process should be very fast. Training on a static model offers the fastest alternative, making it a suitable initial testing ground and offering a good approximation of the system dynamics.
}



\begin{equation}
    \label{eq:static}
\begin{aligned}
    {T_\text{out}}=\frac{1}{qP_\text{cp}+0.4H_\text{l}} \left( \alpha_{K_\text{opt}} n_\text{o} K_\text{opt} I S \right.  \\ - 0.8\left(0.5 T_\text{in}-t_\text{a} \right) (2-\alpha_{H_\text{l}}) \left. +qT_\text{in}P_\text{cp} \right)    
\end{aligned}
\end{equation}

\subsection{Distributed-Parameter Model}

The distributed parameter model describes the energy balances in both the metal and the fluid, incorporating spatially distributed variables~\cite{ControlSolar}. A uniform local concentration ratio is assumed, as the dimensions of the reflector and receiver are considered consistent along the loop, excluding the passive sections. It is assumed that the metal temperature is radially uniform. The loop is discretized longitudinally into 151 segments, each measuring 3.213~m, and computations are performed with an integration time step of 0.25~s to streamline problem resolution. The model is governed by the partial differential Equations~\ref{eq:dpm1} and~\ref{eq:dpm2}.

\begin{equation}
    \label{eq:dpm1}
\begin{aligned}
    \rho_\text{m} C_\text{m} A_\text{m} \frac{\partial T_\text{m}}{\partial t} = n_\text{o} G K_\text{opt} I  \\ + H_\text{l} G (T_\text{a} - T_\text{m}) + LH_\text{t} (T_\text{f} - T_\text{m})
\end{aligned}
\end{equation}

\begin{equation}
\rho_\text{f} C_\text{f} A_\text{f} \frac{\partial T_\text{f}}{\partial t} + q \rho_\text{f} C_\text{f} \frac{\partial T_\text{f}}{\partial x} = - LH_\text{t} (T_\text{f} -T_\text{m})
\label{eq:dpm2}
\end{equation}

\noindent where $A_\text{m}=2.1677\cdot 10^{-4} \text{ m}^2$, $G = 5.75$ m, $L = 0.2136$ , and $A_\text{f} = 0.0036 \text{ m}^2$. 

The convective heat transfer coefficient of the inner tube, $H_\text{t}$, is calculated using Equation~\ref{eq:Ht}~\cite{ControlSolar}.

\begin{equation}\label{eq:Ht}
    \begin{split}
        & H_\text{t} = \left(\frac{q}{3600}\right)^{0.8} \left(-7.182817\cdot10^{-7} T_\text{f}^4 \right. -1.356114\cdot10^3 T_\text{f}^3\\
        &  +2.679214\cdot10^{-1}T_\text{f}^2  +479.1142T_\text{f} + 5.011334\cdot10^3
    \end{split}
\end{equation}

\subsection{Thermal Power}

{The net thermal power, $P_\text{th}$, as described in Equation~\ref{eq:poweri}, is calculated by summing the thermal power contributions from each loop. A penalty term proportional to the flow rate, scaled by a factor $k=3000$, is subtracted. This factor, determined heuristically, discourages excessive adjustments to valve positions and prevents the controller from redirecting all the flow from less efficient loops to the most efficient ones. Such behavior is undesirable as it can lead to increased mechanical wear on the valves, inefficient utilization of the loops, and instability in flow distribution, which negatively impacts system performance.}

\begin{equation}
    \label{eq:poweri}
    P_{\text{th},i} = q_i {\rho_\text{f},i} C_{\text{f},i} (T_{\text{out},i}-T_{\text{in},i}) - kq_i
\end{equation}

The net thermal power is given by equation \ref{eq:power}:

\begin{equation}
    \label{eq:power}
    P_{\text{th}} = \sum_i P_{\text{th},i}
\end{equation}


\section{Proposed Methodology}

\subsection{Defocusing Mechanism and Intercept Factor}

In certain situations, commercial plants may need to adjust one or more collectors by defocusing them, which involves changing their angles to move them out of alignment. This adjustment increases the incidence angle between the solar beam and the normal direction of the mirror plane, reducing the energy captured by the collector and consequently decreasing its efficiency. Defocusing becomes necessary when the outlet temperature exceeds the maximum permissible limit, and increasing the oil flow is not feasible due to constraints in the pump or steam generator. These limitations can arise from energy restrictions or sustained high irradiance levels over a specific period~\cite{MPCdefocusing}.

The defocus curve, as shown in Figure \ref{fig:defocus_curve}, illustrates the relationship between efficiency and defocus angle. Since this relationship is nonlinear, designing a mechanism that can accurately select the appropriate defocus angle is essential. For example, Sánchez et al.~\cite{sanchez2020DefocusingPTC} proposed a method that evaluates defocusing two versus four collectors. The Intercept Factor (IF) is the ratio of solar energy captured by the receiver to the energy reflected by the fraction of incident solar radiation that is effectively intercepted by the receiver~\cite{manikandan2019eff} and considers the effect of defocusing.

\begin{figure}[h!]
    \centering
    \includegraphics[trim=110 10 130 40,clip,width=0.45\textwidth]{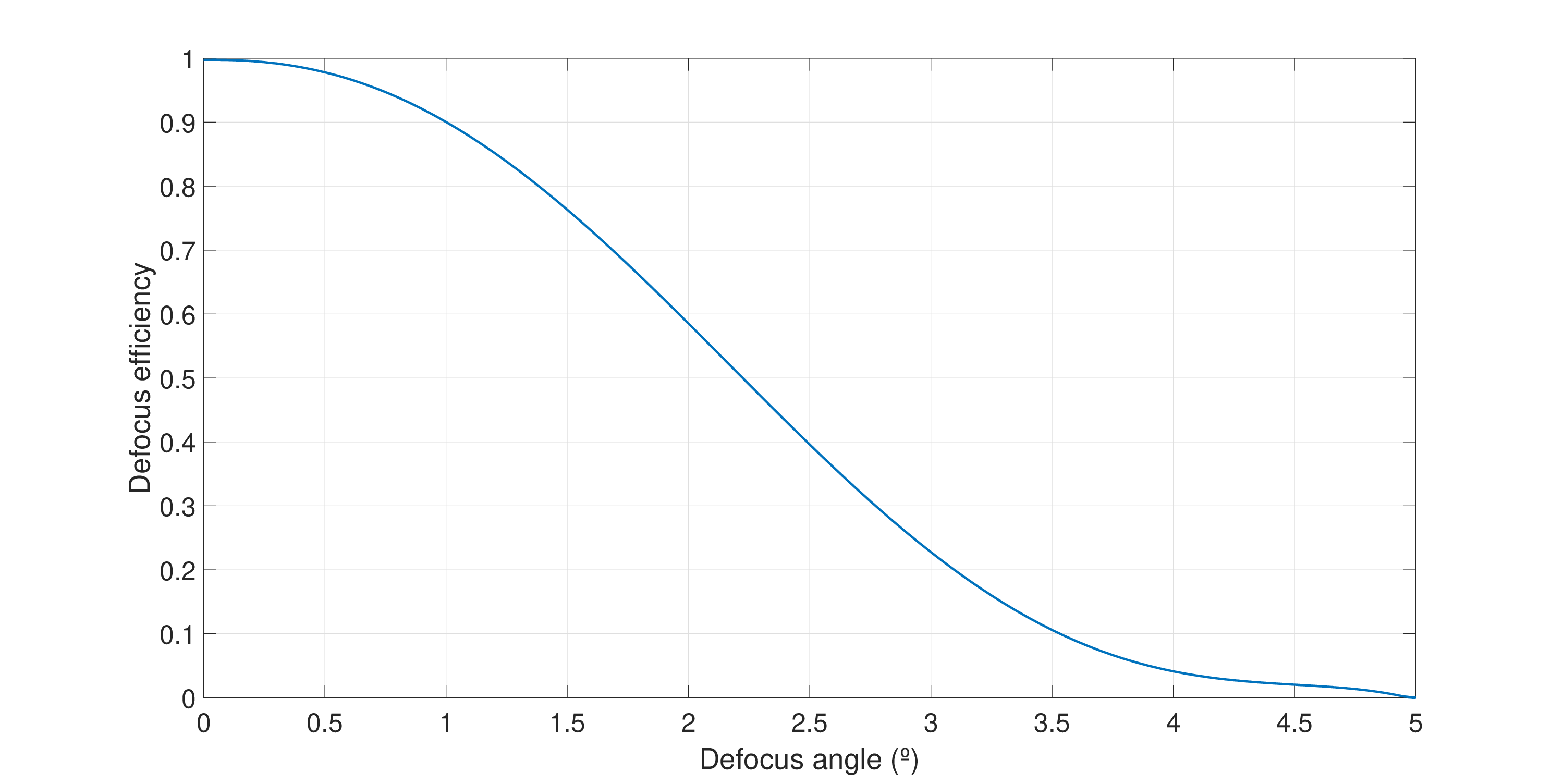}
    \caption{Efficiency-defocus angle curve of the collectors in a PTC plant~\cite{sanchez2020DefocusingPTC}.}
    \label{fig:defocus_curve}
\end{figure}

The first defocusing mechanism was applied to the concentrated parameter model. It involves saturating the outlet temperature whenever it exceeds 392~$^\text{o}$C. The intercept factor is computed as a multiplier to the irradiance in the model of Equation~\ref{eq:concentrated} and is iteratively reduced until the outlet temperature falls below 392~$^\text{o}$C.

For the distributed-parameter model, the defocusing mechanism follows the heuristic method proposed in~\cite{MPCdefocusing}, which is applied to each individual collector. The Intercept Factor (IF) is obtained as the defocusing efficiency. The maximum temperature limits are set as follows:  $T_\text{1, \text{max}}=323^{\text{ o}}$C, $T_\text{2,\text{ max}}=348^{\text{ o}}$C, $T_\text{3,\text{ max}}=373^{\text{ o}}$C, and $T_\text{4,\text{ max}}=390^{\text{ o}}$C. 

\subsection{Auction-Based Methodology}

This work aims to maximize the plant's thermal power while accounting for unknown discrepancies across the loops. To achieve this, a two-layer control scheme for the HTF is implemented. In the first layer, an external controller determines the flow rate for the entire sector, assuming all loops follow the same model. In the second layer, the valves of each loop are locally adjusted using an auction-based methodology to optimize the overall thermal power further. The external controller is assumed to be known and is not the focus of this work. This methodology was applied to the static model from Equation~\ref{eq:static} to ensure fast performance.

To maximize the power output, the auction is computed iteratively for a fixed number of iterations $N_{\text{it}}$. At each iteration, three simulations are conducted for each loop using the static model to predict the resulting power output:

\begin{itemize}
    \item One simulation with the current flow rate $q_i$ for each loop $i$.
    \item One simulation with an increased flow rate $q_i^{+}$ for each loop $i$. 
    \item One simulation with a decreased flow rate $q_i^{-}$ for each looop $i$.
\end{itemize}

The virtual increased and decreased flow rates are given by equation \ref{eq:quantum}, were $\Delta q$ is a pre-selected quantum. The minimum $q_i^{-}$ is saturated to $10^{-6}$~m$^3$\slash s.

\begin{equation}
\label{eq:quantum}
\begin{matrix}
    q_i^{+} &= q_i + \Delta q \\
    q_i^{-} &= q_i - \Delta q
\end{matrix}
\end{equation}

After computing the simulations, the thermal powers are predicted as $P_{\text{th},i}$ using $q_i$, $P_{\text{th},i}^{+}$ using $q_i^{+}$, and $P_{\text{th},i}^{-}$ using $q_i^{-}$. These thermal powers are used to obtain the demand and supply powers as in equation \ref{eq:Pdemsup}:

\begin{equation}
\label{eq:Pdemsup}
\begin{matrix}
    P_{\text{demand},i} &= P_{\text{th},i}^{+} - P_{\text{th},i} \\
    P_{\text{supply},i} &= P_{\text{th},i}^{-} - P_{\text{th},i}
\end{matrix}
\end{equation}

Finally, these powers are used to obtain the supply and demand prices that will be used in the auction. {These are the demand and supply powers scaled by a factor depending on the flow rate increment.}

\begin{equation}
\label{eq:prices}
\begin{matrix}
    C_{\text{demand},i} &= {\frac{1}{\Delta q}}P_{\text{demand},i} \\
    C_{\text{supply},i} &= {\frac{1}{\Delta q}}P_{\text{supply},i}
\end{matrix}
\end{equation}

Then, an auction price $P_\text{au}$ is selected to guide the decision of each loop into demanding, supplying, or maintaining the flow rate. This price can be computed using different criteria: the average power, temperatures, or prices. Based on preliminary experiments aimed at maximizing power and intercept factors, this work computes the auction price as the average of all supply and demand powers as in equation \ref{eq:Pau}:

\begin{equation}
    \label{eq:Pau}
    C_\text{au} = \frac{1}{2N_\text{loops}{\Delta q}} \left( \sum_{i} P_{\text{demand},i} + \sum_{i} P_{\text{supply},i} \right)
\end{equation}

During negotiations, the flow rate of each loop is updated based on three criteria:
\begin{itemize}
    \item To increase the flow rate, the demand price must be greater than the auction price, and the power after increasing must be greater than the power after decreasing and maintaining the flow rate.
    \item To decrease the flow rate, the supply price must be greater than the auction price, and the power after decreasing must be greater after increasing and maintaining the flow rate.
    \item In other cases, the flow rate remains the same.
\end{itemize}

The flow update is proportional to the difference between the supply or demand powers and the auction price. This approach was selected experimentally, although several other criteria were considered, including a constant update, an update proportional to the difference between the auction cost and the powers when increasing or decreasing the flow rates, the difference in powers after maintaining or adjusting the flow rate, and the differences in cost.

Finally, all the flow rates are re-scaled to ensure that the total flow rate of the sector $Q=\sum_i q_i$, as in Equation \ref{eq:reescalado}.

\begin{equation}
\label{eq:reescalado}
    q_i := q_i \frac{Q}{\sum_j q_j}
\end{equation}

Although the algorithm computes the local flow rates, the manipulated variables are the valves' apertures $v_i$, and a transformation must be made at every sample time. Given a valve aperture, the flow rate is obtained with equation \ref{eq:valves}, where the flow rate is computed every $t_{\text{s}1}$ s, and the apertures are updated every $t_{\text{s}2}$ s.

\begin{equation}
\label{eq:valves}
    q_i = Q \frac{v_i}{\sum_j v_j}
\end{equation}

The problem of obtaining the apertures of the valves, given the local flow rates, is under-determined. An iterative process must be made by updating the apertures and computing the flow rates until converging a number of iterations $N_\text{it,v}$. One of the valves is fixed to 100\%, and the rest are updated by a term proportional to the difference between the desired flow rate $q_i$ and the computed at each iteration $\tilde{q}_i$ as in Equation \ref{eq:posvalv}.

\begin{equation}
    \label{eq:posvalv}
    \Delta v = K_\text{v} (q_i - \tilde{q}_i)
\end{equation}

The local control is executed with a $t_{\text{s}2}=3$~min sample time and 10 iterations, and the external control has a sample time of $t_{\text{s}1}=30$~s. The complete process is described in Algorithm \ref{algoritmo} and the parameters chosen by trial and error are $N_{\text{it}}=10$, $N_\text{it,v}=150$, ${\Delta q=1}$ m$^3$\slash s, $K=10^{-5}$, $K_\text{v}=0.25$.

\begin{algorithm} [h]
\label{algoritmo}
\caption{Auction-based flow allocation.}
\label{algoritmo}
\begin{algorithmic}[1]
\State Initialize flow rates as in Equation \ref{eq:valves}
\For{\textbf{each} iteration}
    \For{\textbf{each} loop}
        \State Obtain $q^{+}$ and $q^{-}$ with Equation \ref{eq:quantum}
        \State Predict $P_{\text{th},i}$, $P_{\text{th},i}^{+}$, and $P_{\text{th},i}^{-}$ with Equations~\ref{eq:dpm1} and~\ref{eq:dpm2}
        \State Obtain $P_{\text{demand},i}$, and $P_{\text{supply},i}$ with Equation \ref{eq:Pdemsup}
        \State Obtain $C_{\text{demand},i}$, and $C_{\text{supply},i}$ with Equation \ref{eq:prices}
    \EndFor
    \State Obtain $C_\text{au}$ as in Equation \ref{eq:Pau}
     \For{\textbf{each} loop}
        \If{$C_{\text{demand},i}>C_\text{au}$ and $P_i^{+}>P_i$ and $P_i^{+}>P_i^{-}$}
            \State $q_i := q_i + K(P_{\text{demand},i}-C_\text{au})$
        \ElsIf{$C_{\text{supply},i}>C_\text{au}$ and $P_i^{-}>P_i$ and $P_i^{-}>P_i^{+}$}
            \State $q_i := q_i - K(P_{\text{supply},i}-C_\text{au})$
        \EndIf
    \EndFor   
    \State Saturate flows between $10^{-6}$ m$^3$ \slash s and $Q$ and rescale to ensure the total flow is $Q$ as in Equation \ref{eq:reescalado}
\EndFor
\State Compute valve apertures iteratively with Equation \ref{eq:posvalv}
\end{algorithmic}
\end{algorithm}

\subsection{Artificial Neural Networks}

The objective of this work is to develop a system capable of controlling the actual plant without requiring the iterative algorithm to be computed at each sampling instance. A multilayer perceptron (MLP) was employed to achieve this. An MLP is an ANN consisting of an input layer, an output layer, and one or more hidden layers. Each hidden layer contains a variable number of neurons, with activation functions that introduce nonlinearity.

In this study, the activation functions include a linear function at the output layer, while hyperbolic tangent sigmoid functions are used in the other layers, and the data is scaled within $[-1, 1]$. The network's weights are trained using the Levenberg-Marquardt backpropagation algorithm~\cite{backpropBrain}, which minimizes the sum of squared errors as the loss function. The architecture of the neural networks was determined through a trial-and-error approach. 135 simulations were run using the static model with 27 real irradiance profiles and different values of thermal losses and optical efficiencies to obtain a dataset, and the data were randomized and divided into three subsets: training (70\%), validation (15\%) and test (15\%) sets. 

The input vector to the neural network is $X(k) =(T_\text{in}(k),$ $T_{\text{out},1}(k),$ $T_{\text{out},2}(k), \cdots, T_{\text{out},N_\text{loops}}(k),$ $  T_\text{a} $ $(k),$ $  I(k) n_\text{o}(k), $ $IF_1(k), $ $IF_2(k), \cdots$ $IF_{N_\text{loops}}(k), $ $\bar{T}_\text{out}(k),$ $ \bar{IF}(k),$ $ v_1(k),$ $ v_2(k), \cdots, v_{N_\text{loops}}(k) )$ and the output vector is $Y(k) = (v_1(k+1),$ $ v_2(k+1), \cdots, v_{N_\text{loops}}(k+1) ))$.

\section{Results}

This section shows the results obtained in different simulations. First, the specific characteristics of the methodology, such as the selection of $C_\text{au}$ and the flow rate update, were selected. Next, the entire method was tested by simulation with irradiance profiles that were not included in the training, validation, or test subsets of the neural networks. All computations were performed in \textsc{MATLAB}~R2020b with Intel\textregistered ~Core\texttrademark~ i7-9700F CPU at 3 GHz and 16 GB RAM. Finally, real results of a similar methodology applied to an actual plant are presented.

\begin{table}[t]

\centering
\caption{Average thermal powers and intercept factors of different methodologies in one day simulated with the static model.}
\begin{tabular}{c|c|c}\label{tab:metodos}

{Method}                     & {Power (MW)} & {IF (\%)} \\ \hline 
{No allocation}         & {15.20}                     & {96.07}                    \\ \hline
1 & {15.39}                      & {96.99}                    \\ \hline 
2 & {15.39}                      & {96.97}                    \\ \hline 
3 & {15.39}                      & {96.95}                    \\ \hline 
4 & {15.39}                      & {96.95}                    \\ \hline 
\end{tabular}
\end{table}
\subsection{Simulation Results}

\begin{figure*}[ht] 
    \centering
    \includegraphics[trim=110 20 122 25,clip,width=0.95\linewidth]{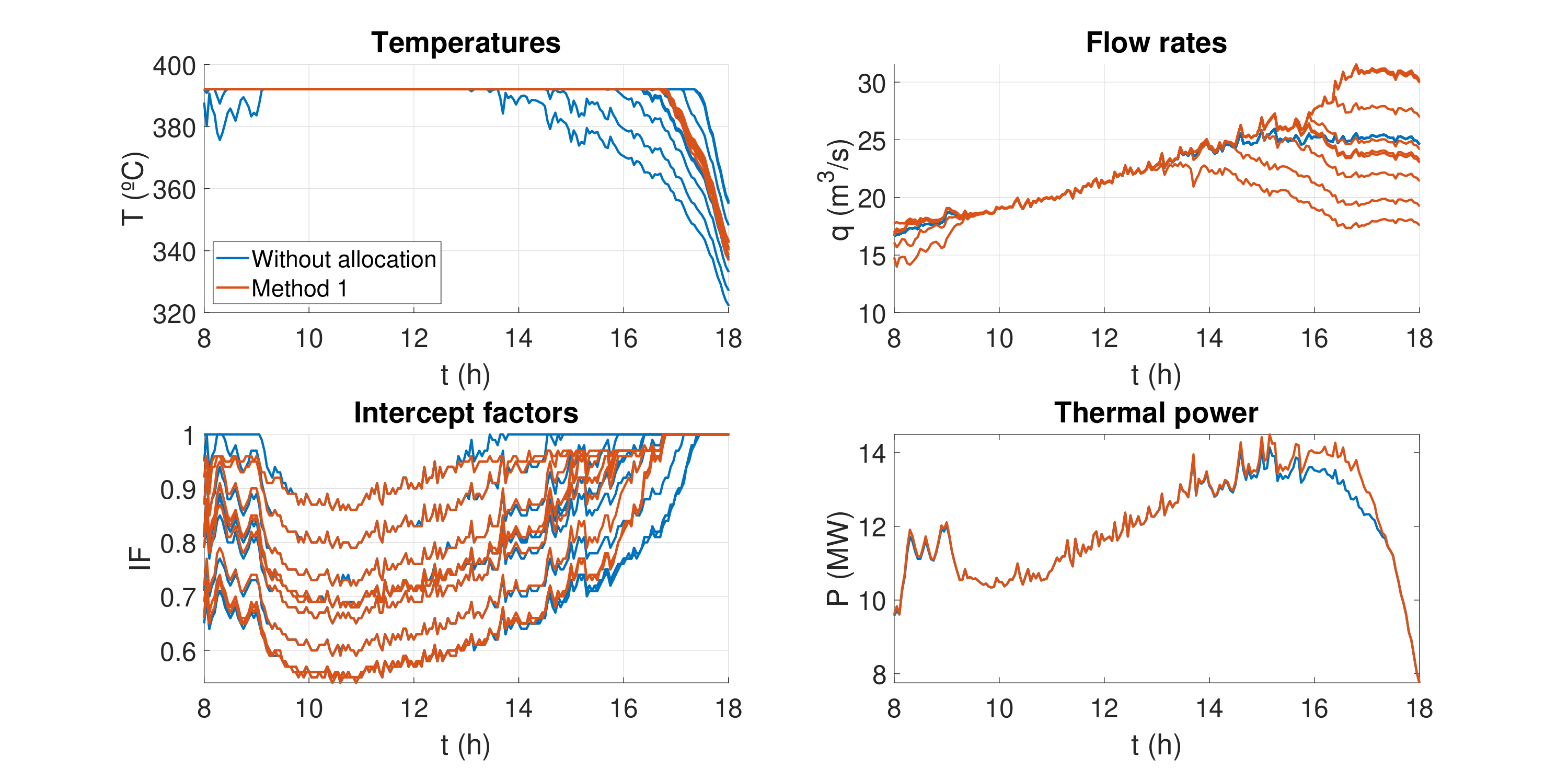}
    \caption{Temperatures, flow rates, intercept factors and thermal powers obtained without allocation and with the first allocation method by simulation of the static model with the first test profile.}\label{fig:estatico1}
\end{figure*}

Table \ref{tab:metodos} depicts the average thermal powers and IF obtained by simulation with the most relevant versions of the methodology with the same irradiance profiles, ambient temperatures, thermal losses, and optical efficiencies in the static model. The case without allocation consists of distributing the flow rate equally to all the loops. The methods shown in the table are: 1) Algorithm \ref{algoritmo}, 2) One auction cost for supply and another one for demand, 3) One auction cost for supply and another one for demand and flow update by $K(P_i^{+}-P_i)$ and $K(P_i^{-}-P_i)$, 4) One auction cost for supply and another one for demand and constant flow update by $K$. Given that the thermal powers with these methods were similar, the selected method (the first method, explained in the previous section) was the one that provided the highest intercept factor.




Three test profiles were used: one cloudy day, one sunny, and one partially cloudy, and the thermal loses and optical efficiencies were randomly selected. Figure \ref{fig:estatico1} compares the temperatures, flow rates, intercept factors, and thermal powers obtained without allocation and with the first method for one of the test irradiance profiles and with the same initial conditions. The method provides a thermal balance in the loops with similar temperatures, even when each loop has different losses and efficiencies. This makes more loops saturate, obtaining higher intercept factors, but the thermal power achieved is higher, especially during the afternoon.

Figure \ref{fig:estatico2} shows the results with the second test profile corresponding to a partially cloudy dataset in the static model. Again, the temperatures are more equally distributed, and the thermal power is higher when the auction-based method is applied. Moreover, during transients, the intercept factors are more similar between each loop.

\begin{figure*}[h]
    \centering
    \includegraphics[trim=110 20 122 25,clip,width=0.95\linewidth]{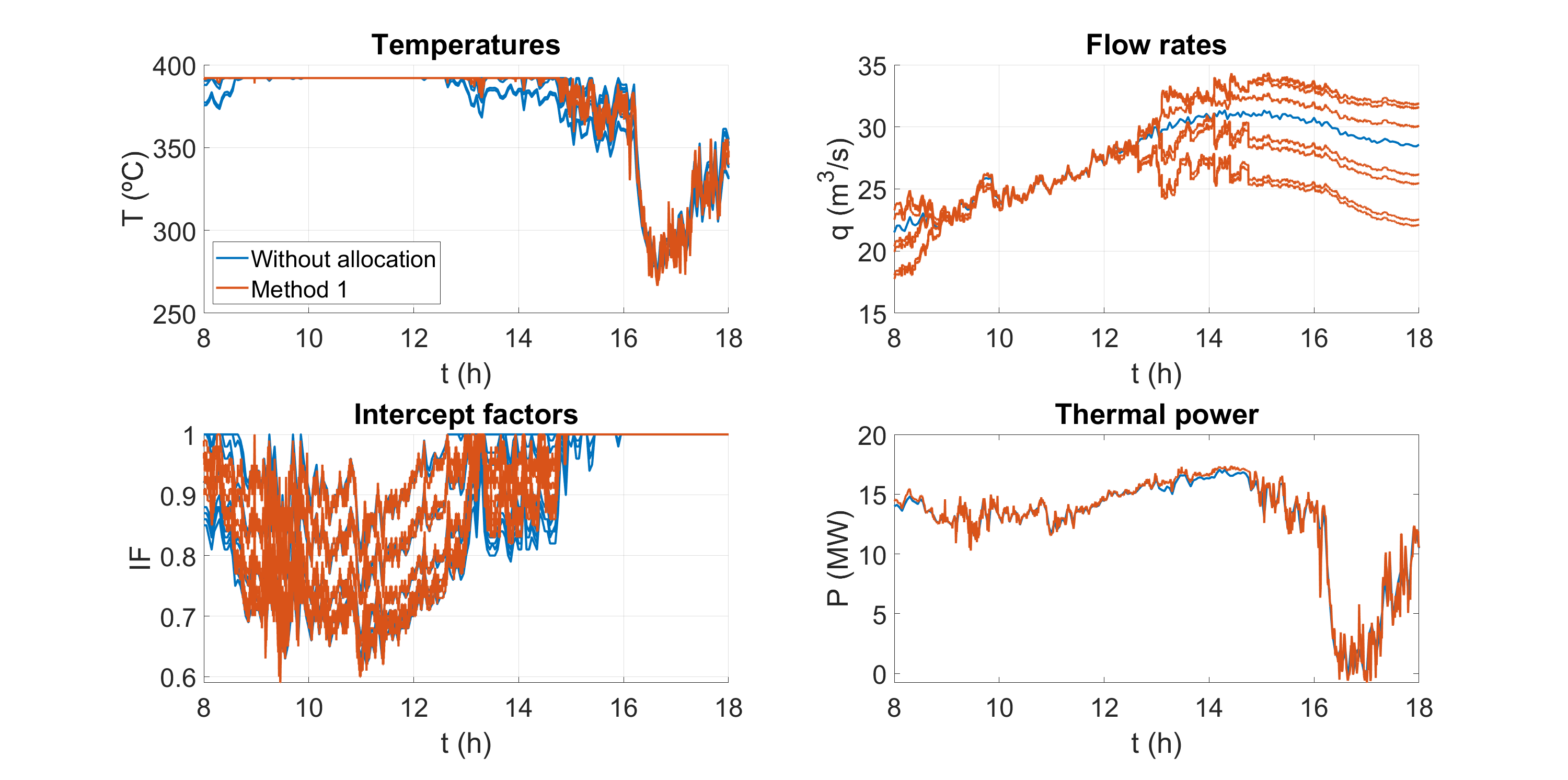}
    \caption{Temperatures, flow rates, intercept factors and thermal powers obtained without allocation and with the first allocation method by simulation of the static model with the second test profile.} \label{fig:estatico2}
\end{figure*}

Finally, results with a cloudy day are shown in Figure \ref{fig:estatico3}. Again, the temperatures are more equally distributed, and the thermal power is higher when applying the methodology. Since the temperatures are not saturating most of the time, the intercept factors are also similar between loops, allowing higher minimal values. 

\begin{figure*}[h!] 
    \centering
    \includegraphics[trim=110 20 122 25,clip,width=0.95\linewidth]{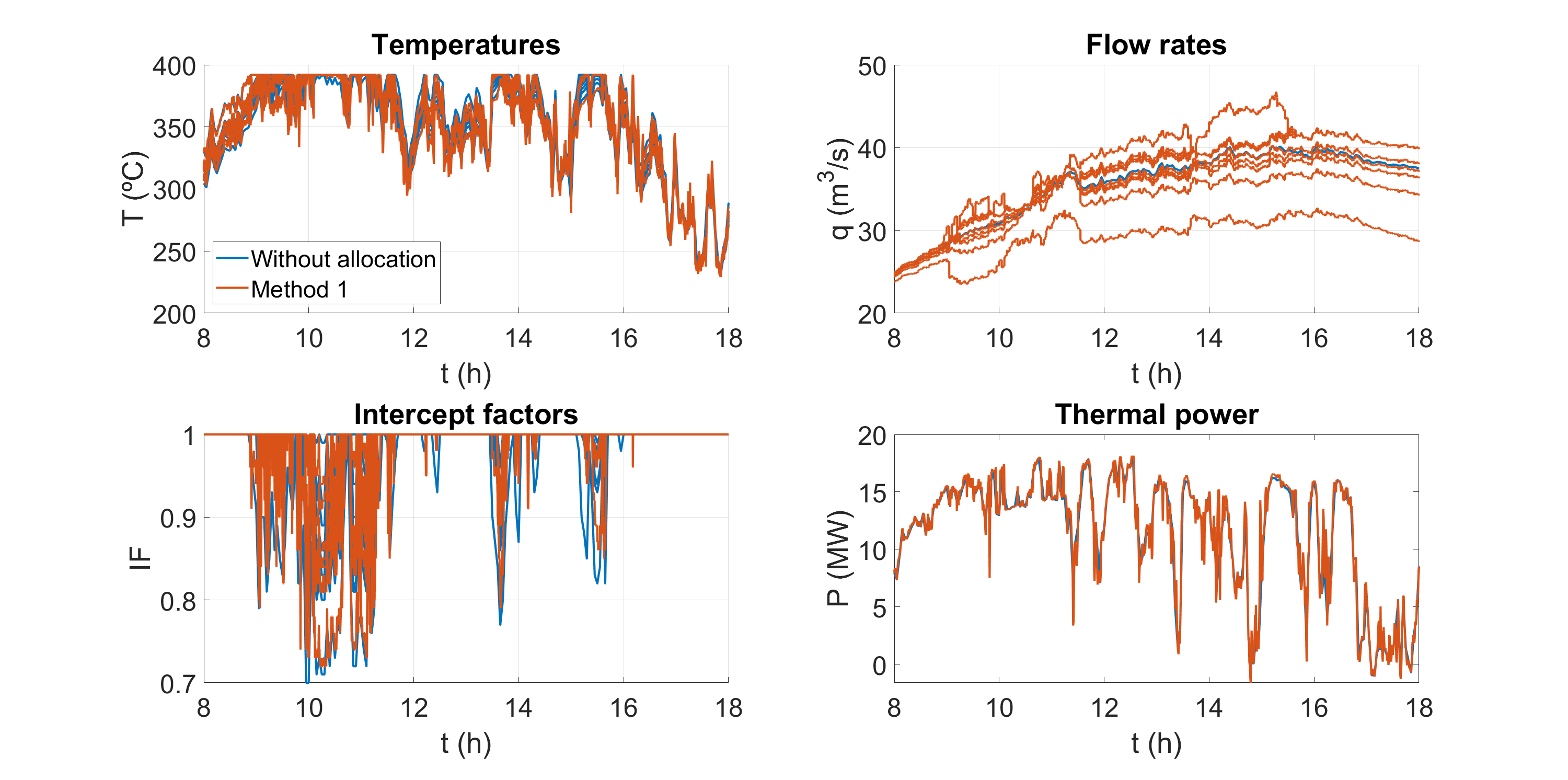}
    \caption{Temperatures, flow rates, intercept factors and thermal powers obtained without allocation and with the first allocation method by simulation of the static model with the third test profile.}\label{fig:estatico3}
\end{figure*}

An extensive training set with different types of weather and operational conditions was generated through simulation. The selected hyper-parameters are given by table \ref{tab:trainingparam}, where $\mu_0$ is the initial damping factor, $\mu$ incr ratio and $\mu$ decr ratio are its increasing and decreasing ratios, max $\mu$ is its maximum value, and max epochs, min gradient and max val checks are the convergence criteria in terms of epochs, gradient and regression checks in the validation subset. Different neural networks with different architectures were trained, and an ANN of three hidden layers with 50, 25 and 10 neurons, respectively, was chosen. The mean squared errors of the valves apertures were $7.59\cdot10^{-6}$ in the training set, $9.57\cdot10^{-6}$ in the validation set, and $1.75\cdot10^{-5}$ in the test set. The correlation coefficients were 99.96\% in the training set, 99.95\% in the validation set, and 99.94\% in the test set{, and the training time was 14.5612 hours}.

\begin{table}[h]
\centering
\caption{Training hyperparameters of the neural networks}
\label{tab:trainingparam}
\resizebox{0.48\textwidth}{!}
{
\begin{tabular}{ccccccc}
 $\mu_0$ & \begin{tabular}[c]{@{}c@{}}$\mu$ incr \\ 
ratio\end{tabular} & \begin{tabular}[c]{@{}c@{}}$\mu$ decr\\ ratio\end{tabular} & \begin{tabular}[c]{@{}c@{}}Max\\ $\mu$\end{tabular} & \begin{tabular}[c]{@{}c@{}}Max\\ epochs\end{tabular} & \begin{tabular}[c]{@{}c@{}}Min\\ gradient\end{tabular} & \begin{tabular}[c]{@{}c@{}}Max\\ val checks\end{tabular} \\ \hline
 $10^{-3}$ & $10 $                    & $10^{-1}$   & $10^{10}$  & $4\cdot10^3$   & $10^{-7}$   & 6  \\ 
\end{tabular}
}
\end{table}

The neural networks were tested first with the concentrated-parameter model to analyze their behavior. Table \ref{tab:res_power} shows the thermal powers of simulations with the three test profiles and their average values. In addition, a ponderated average is shown based on the average portion of sunny days (57.5\%), partially cloudy days (42.28\%), and cloudy days (0.22\%) in Gila Bend. The impact factors are shown in Table \ref{tab:res_if}. Both allocation strategies improve the results with respect to distributing the flow equally in terms of thermal power and intercept factors. The average compuation times were $1.03\cdot10^{-3}$ s with the allocation method and $6.88\cdot10^{-5}$ with the neural networks, two orders of magnitude shorter.

\begin{table}[h]\label{tab:res_power}

\centering
\caption{Thermal powers (kW) obtained without flow allocation, with the first allocation method and with the ANN by simulating the static model with the three test profiles and their mean and weighted mean values.}
\resizebox{0.48\textwidth}{!}
{\begin{tabular}{c|c|c|c|c|c}\label{tab:res_power}
Method        & Test 1 & Test 2 & Test 3 & Mean  & W. mean \\ \hline
No allocation & {11.90}   & {12.63}  & {11.50}   & {12.01}  & {12.21}          \\ \hline
Method 1      & {12.04}   & {12.76}  & {11.63}   & {12.14}  & {12.34}          \\ \hline
ANN           & {12.04}   & {12.72}  & {11.55}   & {12.10}  & {12.33}          \\ \hline
\end{tabular}}
\end{table}

\begin{table}[h]

\centering
\caption{Intercept factors (\%) obtained without flow allocation, with the first allocation method and with the ANN by simulating the static model with the three test profiles and their mean values.}
\resizebox{0.48\textwidth}{!}
{\begin{tabular}{c|c|c|c|c|c}\label{tab:res_if}
Method        & Test 1 & Test 2 & Test 3 & Mean  & W. mean \\ \hline
No allocation & 79.72 & 88.80 & 98.17 & 88.90 & 83.60         \\ \hline
Method 1      & 80.32 & 89.12 & 98.43 & 89.29 & 84.08         \\ \hline
ANN           & 80.30 & 89.14 & 98.39 & 89.28 & 84.08         \\ \hline
\end{tabular}}
\end{table}

{
\begin{figure*}[h!] 
    \centering
    \includegraphics[trim=110 20 122 25,clip,width=0.95\linewidth]{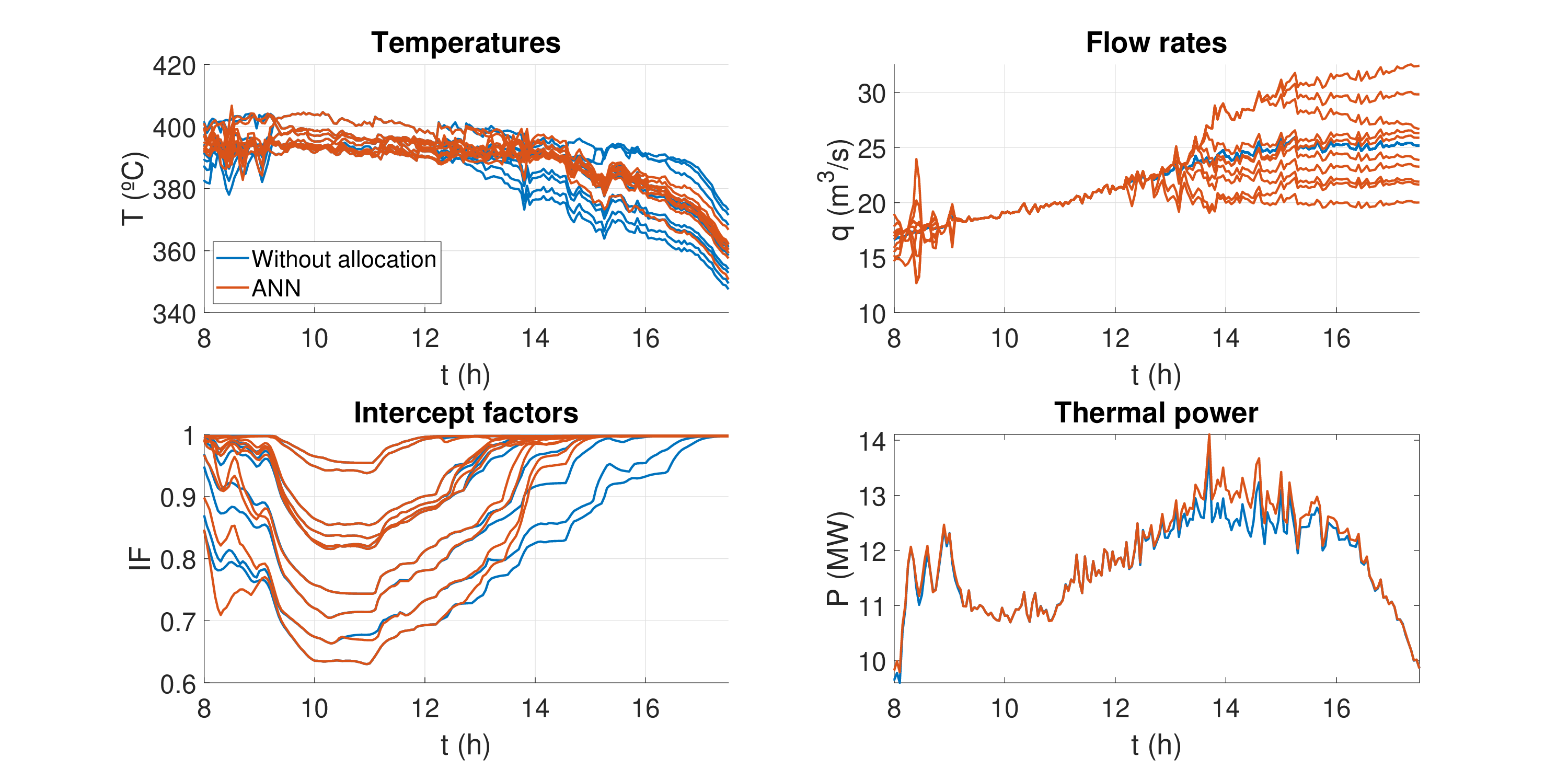}
    \caption{Temperatures, flow rates, intercept factors and thermal powers obtained without allocation and with the ANN by simulation of the distributed-parameter model with the first test profile.}
    \label{fig:distrib1}
\end{figure*}

\begin{figure*}[h!]
    \centering
    \includegraphics[trim=110 20 122 25,clip,width=0.95\linewidth]{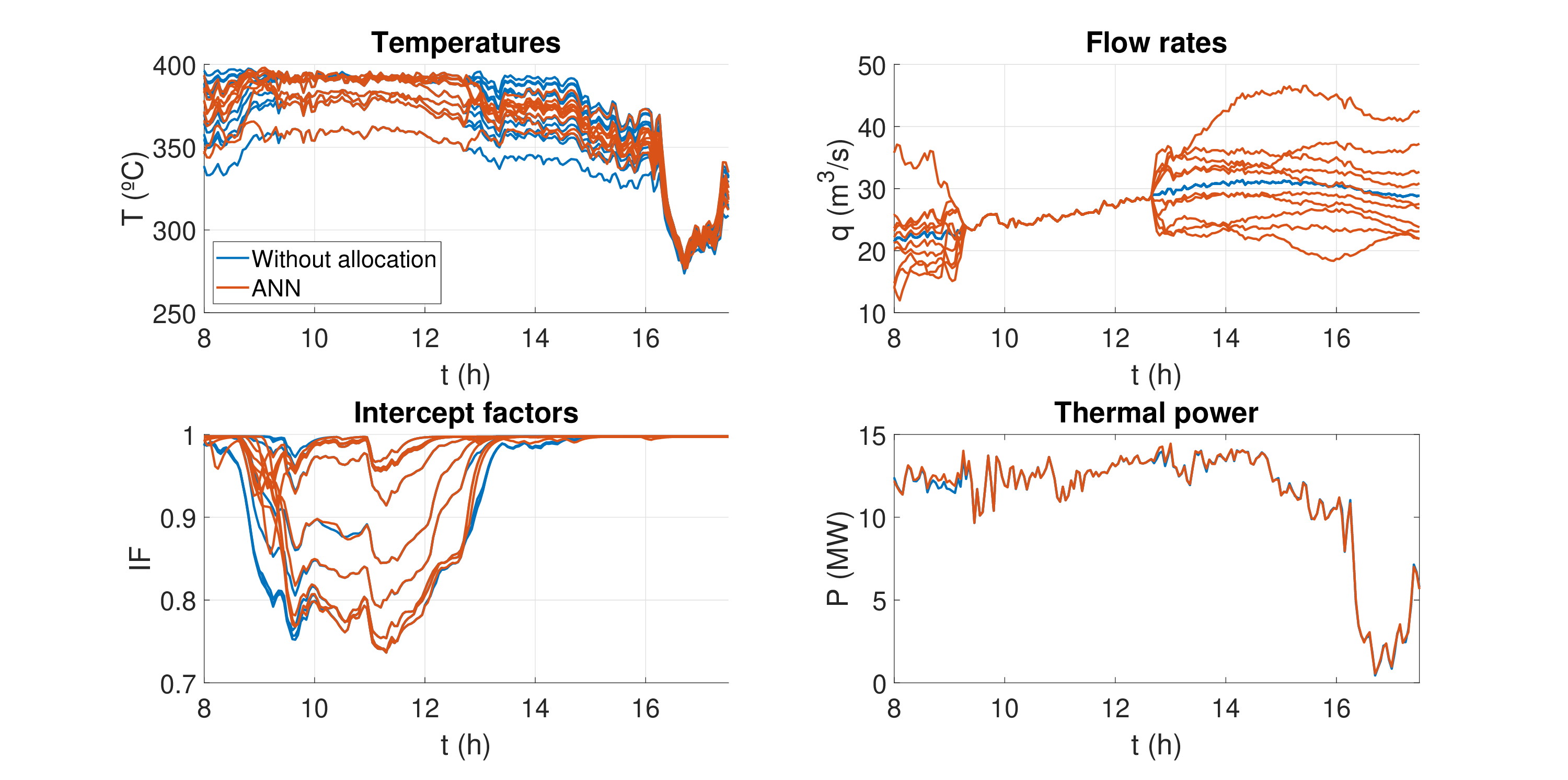}
    \caption{Temperatures, flow rates, intercept factors and thermal powers obtained without allocation and with the ANN by simulation of the distributed-parameter model with the second test profile.}
 \label{fig:distrib2}
\end{figure*}

}

\begin{figure*}[h] 
    \centering
    \includegraphics[trim=110 20 122 25,clip,width=0.95\linewidth]{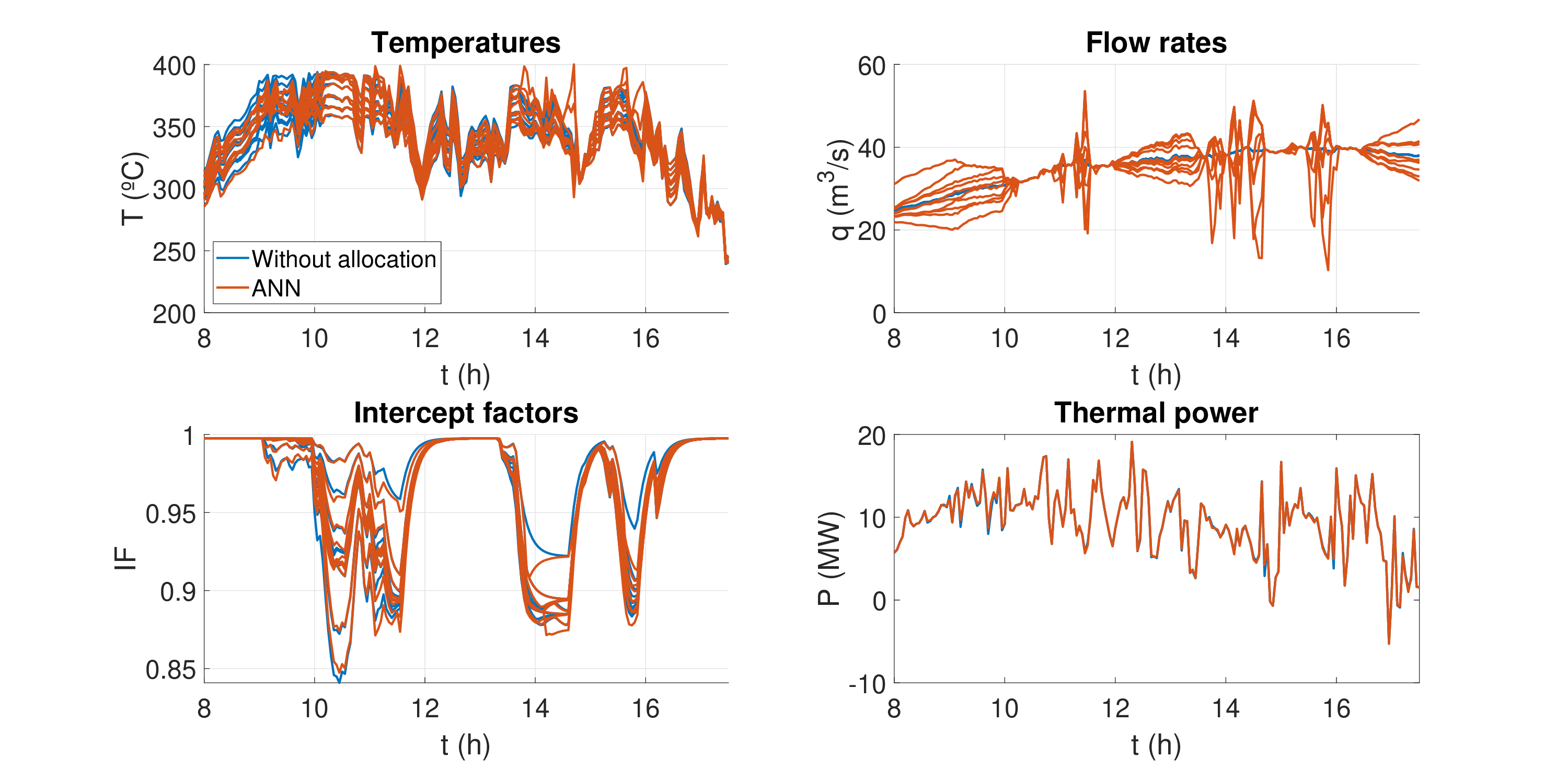}
    \caption{Temperatures, flow rates, intercept factors and thermal powers obtained without allocation and with the ANN by simulation of the distributed-parameter model with the third test profile.}
    \label{fig:distrib3}
\end{figure*}

The same irradiance profiles and simulation conditions were used to test the method with the distributed-parameter model. In this case, the intercept factors were filtered with a low pass filter of 10 min of time constant. Tables \ref{tab:res_power_dist} and \ref{tab:res_if_dis} show the results without flow allocation and with the neural network. The basic algorithm without neural networks has not been tested on this model, as it would be necessary to integrate the distributed-parameter model over hours to obtain the steady-state response at each sampling time. Nevertheless, these simulations demonstrate the adaptability of neural controllers to a more complex system. The tests with this model show similar improvements as with the static model, obtaining better thermal powers and intercept factors after applying the proposed methodology. {The average computation time of the ANN was $1.35\cdot10^{-4}$~s. }


\begin{table}[h!]

\centering
\caption{Thermal powers (kW) obtained without flow allocation and with the ANN by simulating the distributed-parameter model with the three test profiles and their mean and weighted mean values.}
\resizebox{0.48\textwidth}{!}
{\begin{tabular}{c|c|c|c|c|c}\label{tab:res_power_dist}
Method        & Test 1 & Test 2 & Test 3 & Mean  & W. mean \\ \hline
No allocation & {11.75} & {11.20} & { 9.40} & {10.78} & {11.51}         \\ \hline
ANN           & {11.86} & {11.25} & { 9.43} & {10.85} & {11.60}        \\ \hline
\end{tabular}}
\end{table}

\begin{table}[h!]

\centering
\caption{Intercept factors (\%) obtained without flow allocation and with the ANN by simulating the distributed-parameter model with the three test profiles and their mean values.}
\resizebox{0.48\textwidth}{!}
{\begin{tabular}{c|c|c|c|c|c}\label{tab:res_if_dis}
Method        & Test 1 & Test 2 & Test 3 & Mean  & W. mean \\ \hline
No allocation & 91.22 & 95.69 & 96.65 & 94.52 & 93.12        \\ \hline
ANN           & 91.97 & 95.98 & 96.62 & 94.85 & 93.67   \\ \hline
\end{tabular}}
\end{table}

Figures \ref{fig:distrib1}, \ref{fig:distrib2} and \ref{fig:distrib3} show the results with the distributed-parameter model. Although in both of them, the thermal powers are higher and the temperatures are more balanced, this is more visible in the first test, with greater improvements the clearer the day.

\subsection{Application to Real Plants}

A controller based in the previous algorithms and adapted to the plant distributed control system (DCS) has been tested and} applied to 13 of Atlantica Sustainable Infrastructure Ltd's 50 MW solar trough plants. The control algorithms installed on the DCS has to fulfill two requirements: i) the computational requirements have to be as low as possible and ii) the algorithms cannot use a complex optimization library solver. Thus the approach proposed in this paper is well suited to be applied on these plants. The controller was initially tested in the Helionergy 1 plant, located near Écija in Southern Spain. Results corresponding to this test are shown below.

The plant's solar field aperture area is {about} 300,000~m$^2$, with 360 Solar Collector Assemblies (SCAs) structured in 90 loops, with 4 SCAs per loop. The SCA length is 150~m, and the collector model used is the Astro (ET-150). The nominal turbine power is 50~MW, operating at 100 bars.

\begin{figure*}[h!] 
    \centering
    \includegraphics[width=0.7\linewidth]{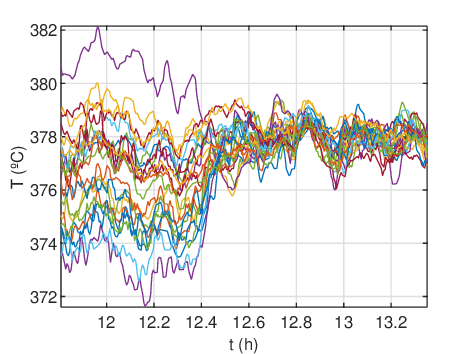}
    \caption{Temperatures obtained in the Helionergy 1 plant. The controller was activated after $t=12.4$~h.}
    \label{fig:Res_03_e}
\end{figure*}

Figure \ref{fig:Res_03_e} shows the evolution of temperatures in the west quadrant of the plant before and after the controller was activated. Before activation, the temperature fluctuated within an 8~$^\text{o}$C range, as all the loops were receiving the same HTF flow, with the most efficient loops reaching higher temperatures. Once the controller was activated, HTF was exchanged between loops, transferring HTF from the less efficient loops to the more efficient ones. More energy was collected in the solar field because the gain in collected energy in the more efficient loops outweighed the energy lost in the less efficient loops due to the reduced HTF they received. Furthermore, since the higher temperatures decreased while maintaining the solar field's average temperature, the number of defocusing operations decreased. This reduced energy losses due to defocusing and will also be beneficial in the long term by lowering maintenance costs, as the collector actuators will be used less.

After a test campaign at the Helionergy plant, the controller was commissioned in 13 of Atlantica Renewable Infrastructure's 50~MW plants.

\section{Conclusions}

This paper presents a novel approach to optimizing the thermal balance in PTC plants using a market-based system for flow distribution among loops combined with an ANN to reduce computational costs and the amount of data needed, making it suitable to be applied in an actual plant, where the computers cannot compute optimization problems or complex algorithms. The methodology was tested under various conditions, showing promising results. The proposed auction-based method effectively balances temperatures across different loops, even when loops have varying thermal losses and efficiencies. The methodology was tested using different irradiance profiles (sunny, partially cloudy, and cloudy days) and consistently outperformed the no-allocation baseline in terms of both thermal power and intercept factors. The results indicate that the method is robust and adaptable to varying environmental conditions. The adaptation from simulation with the concentrated-parameter model to the distributed-parameter model underscores the generalizability of the method across different modeling paradigms. The reduction in the information needed by the flow allocation achieved through the ANN makes this approac h scalable and implementable in real-world scenarios, potentially leading to significant operational efficiency improvements in large-scale solar thermal plants.

Feature work could focus on refining the ANN architectures and the optimization methodology to improve performance and adaptability to more complex plant configurations and additional environmental variables. 

\section*{Declaration of Competing Interest}
The authors declare that they have no known competing financial interests or personal relationships that could have appeared to influence the work reported in this paper.

\section*{Acknowledgement}
This paper has received funding from the European Research Council (ERC) under the Proof of Concept Lump Sum Grant (OCoSEP, grant agreement No~ERC-2022-PoC1), and by the Spanish Ministry of Science, Innovation and Universities (FPU, grant No~FPU20/01958, and Control Coalicional para la Optimización de Sistemas Ciberfísicos: Ronda 3, grant No~ PID2023-152876OB-I00).

The authors would like to thank Elizabeth F. Buckley for her assistance in adapting and programming the algorithms for deployment in real solar plants.

\section{Data Availability}

The dataset used in this work is publicly available in \url{https://zenodo.org/records/14645687}.

\bibliography{references.bib}

\begin{thebibliography}{10}
\expandafter\ifx\csname url\endcsname\relax
  \def\url#1{\texttt{#1}}\fi
\expandafter\ifx\csname urlprefix\endcsname\relax\def\urlprefix{URL }\fi
\expandafter\ifx\csname href\endcsname\relax
  \def\href#1#2{#2} \def\path#1{#1}\fi

\bibitem{solarTrends}
Z.~Şen, Solar energy in progress and future research trends, Progress in Energy and Combustion Science 30 (2004) 367 -- 416.
\newblock \href {http://dx.doi.org/https://doi.org/10.1016/j.pecs.2004.02.004} {\path{doi:https://doi.org/10.1016/j.pecs.2004.02.004}}.

\bibitem{PVsurvey}
E.~Rakhshani, K.~Rouzbehi, A.~J. Sánchez, A.~C. Tobar, E.~Pouresmaeil, Integration of large scale \{PV\}-based generation into power systems: A survey, Energies 12 (2019) 1425.
\newblock \href {http://dx.doi.org/https://doi.org/10.3390/en12081425} {\path{doi:https://doi.org/10.3390/en12081425}}.

\bibitem{wec_humanising}
World energy council. humanising energy: a look at the {G20} agenda (12 2024).

\bibitem{ahmad2023PTC}
A.~Ahmad, O.~Prakash, R.~Kuasher, G.~Kumar, S.~Pandey, S.~M.~M. Hasnain, Parabolic trough solar collectors: A sustainable and efficient energy source, Materials Science for Energy Technologies\href {http://dx.doi.org/https://doi.org/10.1016/j.mset.2023.08.002} {\path{doi:https://doi.org/10.1016/j.mset.2023.08.002}}.

\bibitem{NMPCbalance}
A.~J. Gallego, A.~J. Sánchez, J.~M. Escaño, E.~F. Camacho, Nonlinear model predictive control for thermal balance in solar trough plants, European Journal of Control 67 (2022) 100717.
\newblock \href {http://dx.doi.org/https://doi.org/10.1016/j.ejcon.2022.100717} {\path{doi:https://doi.org/10.1016/j.ejcon.2022.100717}}.

\bibitem{pyrheliometer}
Pirheliometer {DR03} (12 2021).

\bibitem{song2024homoKalman}
Y.~Song, J.~Wang, J.~Zhang, Y.~Li, Temperature homogenization control of parabolic trough solar collector field based on hydraulic calculation and extended kalman filter, Renewable Energy (2024) 120346.

\bibitem{camacho2024control}
E.~F. Camacho, S.~Ruiz-Moreno, J.~M. Aguila-López, A.~J. Gallego, R.~A. García, Control of solar energy systems, Annual Review of Control, Robotics, and Autonomous Systems 7~(Volume 7, 2024) (2024) 175--200.
\newblock \href {http://dx.doi.org/https://doi.org/10.1146/annurev-control-071023-103936} {\path{doi:https://doi.org/10.1146/annurev-control-071023-103936}}.

\bibitem{TempHomo}
A.~J. Sánchez, A.~J. Gallego, J.~M. Escaño, E.~F. Camacho, Temperature homogenization of a solar trough field for performance improvement, Solar Energy 165 (2018) 1 -- 9.
\newblock \href {http://dx.doi.org/https://doi.org/10.1016/j.solener.2018.03.001} {\path{doi:https://doi.org/10.1016/j.solener.2018.03.001}}.

\bibitem{ThermalBalance}
A.~J. Sánchez, A.~J. Gallego, J.~M. Escaño, E.~F. Camacho, Thermal balance of large scale parabolic trough plants: A case study, Solar Energy 190 (2019) 69 -- 81.
\newblock \href {http://dx.doi.org/https://doi.org/10.1016/j.solener.2019.08.001} {\path{doi:https://doi.org/10.1016/j.solener.2019.08.001}}.

\bibitem{gallego2023tcp100}
A.~J. Gallego, L.~J. Yebra, A.~J. S.~D. Pozo, J.~M. Escaño, E.~F. Camacho, Nonlinear mpc for thermal balancing of the {TCP-100} parabolic trough collectors solar plant, 2023, pp. 1807--1812.
\newblock \href {http://dx.doi.org/https://doi.org/10.23919/ACC55779.2023.10156440} {\path{doi:https://doi.org/10.23919/ACC55779.2023.10156440}}.

\bibitem{frejo2020CentralDistributedMPC}
J.~R.~D. Frejo, E.~F. Camacho, Centralized and distributed model predictive control for the maximization of the thermal power of solar parabolic-trough plants, Solar Energy 204 (2020) 190--199.
\newblock \href {http://dx.doi.org/10.1016/j.solener.2020.04.033} {\path{doi:10.1016/j.solener.2020.04.033}}.

\bibitem{chanfreut2023clustering}
P.~C. Palacio, J.~M. Maestre, A.~J. Gallego, A.~Annaswamy, E.~F. Camacho, Clustering-based model predictive control of solar parabolic trough plants, Jose M. and Gallego, Antonio J. and Annaswamy, Anuradha and Camacho, Eduardo F., Clustering-Based Model Predictive Control of Solar Parabolic Trough Plants\href {http://dx.doi.org/https://doi.org/10.1016/j.renene.2023.118978} {\path{doi:https://doi.org/10.1016/j.renene.2023.118978}}.

\bibitem{sanchez2023populationDynamics}
A.~Sánchez-Amores, J.~Martinez-Piazuelo, J.~M. Maestre, C.~Ocampo-Martinez, E.~F. Camacho, N.~Quijano, Population-dynamics-assisted coalitional model predictive control for parabolic-trough solar plants, IFAC-PapersOnLine 56 (2023) 7710--7715.
\newblock \href {http://dx.doi.org/https://doi.org/10.1016/j.ifacol.2023.10.1174} {\path{doi:https://doi.org/10.1016/j.ifacol.2023.10.1174}}.

\bibitem{schimperna2024rnnmpc}
I.~Schimperna, G.~Galuppini, L.~Magni, Recurrent neural network based mpc for systems with input and incremental input constraints, IEEE Control Systems Letters\href {http://dx.doi.org/https://doi.org/10.1109/LCSYS.2024.3404332} {\path{doi:https://doi.org/10.1109/LCSYS.2024.3404332}}.

\bibitem{hose2023approxMPC}
H.~Hose, J.~Köhler, M.~N. Zeilinger, S.~Trimpe, Approximate non-linear model predictive control with safety-augmented neural networks, arXiv preprint\href {http://dx.doi.org/https://doi.org/10.48550/arXiv.2304.09575} {\path{doi:https://doi.org/10.48550/arXiv.2304.09575}}.

\bibitem{ANNi_PTC}
M.~Cervantes-Bobadilla, J.~A. Hernández-Pérez, D.~Juárez-Romero, A.~Bassam, J.~García-Morales, A.~Huicochea, O.~A. Jaramillo, Control scheme formulation for a parabolic trough collector using inverse artificial neural networks and particle swarm optimization, Journal of the Brazilian Society of Mechanical Sciences and Engineering 43 (2021) 1--14.
\newblock \href {http://dx.doi.org/https://doi.org/10.1007/s40430-021-02862-4} {\path{doi:https://doi.org/10.1007/s40430-021-02862-4}}.

\bibitem{goel2021genetic}
A.~Goel, O.~P. Verma, G.~Manik, Flow rate optimization of a parabolic trough solar collector using multi-objective genetic algorithm (2022).

\bibitem{tilahun2024fuzzy}
F.~B. Tilahun, Fuzzy-based predictive deep reinforcement learning for robust and constrained optimal control of industrial solar thermal plants, Applied Soft Computing 159 (2024) 111432.
\newblock \href {http://dx.doi.org/https://doi.org/10.1016/j.asoc.2024.111432} {\path{doi:https://doi.org/10.1016/j.asoc.2024.111432}}.

\bibitem{masero2022marketMPC}
E.~Masero, J.~M. Maestre, E.~F. Camacho, Market-based clustering of model predictive controllers for maximizing collected energy by parabolic-trough solar collector fields, Applied Energy 306 (2022) 117936.
\newblock \href {http://dx.doi.org/https://doi.org/10.1016/j.apenergy.2021.117936} {\path{doi:https://doi.org/10.1016/j.apenergy.2021.117936}}.

\bibitem{modelMojave}
A.~J. Gallego, M.~Macías, F.~de~Castilla, E.~F. Camacho, Mathematical modeling of the \{Mojave\} solar plants, Energies 12 (2019) 4197.
\newblock \href {http://dx.doi.org/https://doi.org/10.3390/en12214197} {\path{doi:https://doi.org/10.3390/en12214197}}.

\bibitem{therminolvp1}
\href{https://www.therminol.com/product/71093459?pn=Therminol-VP-1-Heat-Transfer-Fluid}{Therminol vp1 htf, 2023}.
\newline\urlprefix\url{https://www.therminol.com/product/71093459?pn=Therminol-VP-1-Heat-Transfer-Fluid}

\bibitem{MPCmojave}
A.~J. Gallego, M.~Macías, F.~de~Castilla, A.~J. Sánchez, E.~F. Camacho, Model predictive control of the \{Mojave\} solar trough plants, Control Engineering Practice 123 (2022) 105140.
\newblock \href {http://dx.doi.org/https://doi.org/10.1016/j.conengprac.2022.105140} {\path{doi:https://doi.org/10.1016/j.conengprac.2022.105140}}.

\bibitem{princsolar}
D.~Y. Goswami, F.~Kreith, J.~F. Kreider, Principles of solar engineering, CRC Press, 2000.
\newblock \href {http://dx.doi.org/https://doi.org/10.1201/9781003244387} {\path{doi:https://doi.org/10.1201/9781003244387}}.

\bibitem{modelTCP100}
A.~J.~G. Len, L.~J. Yebra, E.~F. Camacho, A.~J. Sánchez, Mathematical modeling of the parabolic trough collector field of the \{TCP-100\} research plant, Linköping University Electronic Press, Linköpings universitet, 2018, pp. 912--918.
\newblock \href {http://dx.doi.org/https://doi.org/10.3384/ecp17142912} {\path{doi:https://doi.org/10.3384/ecp17142912}}.

\bibitem{dynamicModelPTC}
R.~Österholm, J.~Pålsson, Dynamic modelling of a parabolic trough solar power plant, 2014, pp. 409--418.
\newblock \href {http://dx.doi.org/http://dx.doi.org/10.3384/ecp140961057} {\path{doi:http://dx.doi.org/10.3384/ecp140961057}}.

\bibitem{ControlSolar}
E.~F. Camacho, M.~Berenguel, F.~R. Rubio, D.~Martinez, Control of solar energy systems, Springer-Verlag, 2012.
\newblock \href {http://dx.doi.org/https://doi.org/10.1007/978-0-85729-916-1} {\path{doi:https://doi.org/10.1007/978-0-85729-916-1}}.

\bibitem{MPCdefocusing}
A.~J. Sánchez, A.~J. Gallego, J.~M. Escaño, E.~F. Camacho, Event-based mpc for defocusing and power production of a parabolic trough plant under power limitation, Solar Energy 174 (2018) 570 -- 581.
\newblock \href {http://dx.doi.org/https://doi.org/10.1016/j.solener.2018.09.044} {\path{doi:https://doi.org/10.1016/j.solener.2018.09.044}}.

\bibitem{sanchez2020DefocusingPTC}
A.~J. Sánchez, A.~J. Gallego, J.~M. Escaño, E.~F. Camacho, Parabolic trough collector defocusing analysis: Two control stages vs four control stages, Solar Energy 209 (2020) 30 -- 41.
\newblock \href {http://dx.doi.org/https://doi.org/10.1016/j.solener.2020.09.001} {\path{doi:https://doi.org/10.1016/j.solener.2020.09.001}}.

\bibitem{manikandan2019eff}
G.~K. Manikandan, S.~Iniyan, R.~Goic, \href{https://www.sciencedirect.com/science/article/pii/S0306261918317537}{Enhancing the optical and thermal efficiency of a parabolic trough collector – a review}, Applied Energy 235 (2019) 1524--1540.
\newblock \href {http://dx.doi.org/https://doi.org/10.1016/j.apenergy.2018.11.048} {\path{doi:https://doi.org/10.1016/j.apenergy.2018.11.048}}.
\newline\urlprefix\url{https://www.sciencedirect.com/science/article/pii/S0306261918317537}

\bibitem{backpropBrain}
T.~P. Lillicrap, A.~Santoro, L.~Marris, C.~J. Akerman, G.~Hinton, Backpropagation and the brain, Nature Reviews Neuroscience 21 (2020) 335--346.
\newblock \href {http://dx.doi.org/https://doi.org/10.1038/s41583-020-0277-3} {\path{doi:https://doi.org/10.1038/s41583-020-0277-3}}.

\end{thebibliography}

\end{document}